\documentstyle[10pt,epic,eepic]{siam}

\newenvironment{tabAlgorithm}[2]{
\setcounter{algorithmLine}{1}
\samepage
\begin{tabbing}
999\=\kill
#1 \ --- \ {\it #2}
}{
\end{tabbing}
}
\newcounter{algorithmLine}
\newcommand{\algline}{\\\thealgorithmLine\hfil\>\stepcounter{algorithmLine}}
 
\def\myendproof{\hfill{\vbox{\hrule\hbox{%
   \vrule height1.3ex\hskip0.8ex\vrule}\hrule}}}

\newcommand{\fig}[3] 
{\begin{figure}[htbp]
 \begin{center}
 \input{#1}
 \end{center}
 \caption{#3}
 \label{#2}
 \end{figure}
}

\newcommand{\Dim}[1]{\Re^{#1}}
\newcommand{\DimD}{\Dim{d}}
\newcommand{\Perimeter}[1]{\overline{#1}}
\newcommand{\Edge}[1]{\overline{#1}}

\newcommand{\Angle}[1]{\angle #1}

\title{Low Degree Spanning Trees of Small Weight}

\author{Samir Khuller 
\thanks{Department of Computer Science and Institute for Advanced 
Computer Studies, University of Maryland, College Park, MD~20742. 
Research supported by NSF Research Initiation Award CCR-9307462.
E-mail : {\tt samir@cs.umd.edu}.}
\and Balaji Raghavachari
\thanks{Department of Computer Science, The University of Texas at Dallas,
Richardson, TX 75083.
Research supported by NSF Research Initiation Award CCR-9409625.
E-mail : {\tt rbk@utdallas.edu}.}
\and Neal Young
\thanks{Department of Computer Science,
Dartmouth College, Hanover, NH 03755-3510.
This work was done while the author was at Cornell University and at UMIACS
and was supported in part by NSF grants CCR-8906949 and CCR-9111348.
E-mail : {\tt neal.young@dartmouth.edu}.}
}

\begin{document}

\maketitle
\thispagestyle{empty}

\begin{abstract}
Given $n$ points in the plane, the degree-$K$ spanning tree problem
asks for a spanning tree of minimum weight in which the degree of each
vertex is at most $K$. 
This paper addresses the problem of computing low-weight degree-$K$
spanning trees for $K>2$. It is shown that for an arbitrary
collection of $n$ points in the plane, there exists a spanning tree of
degree three whose weight is at most 1.5 times the weight of a minimum
spanning tree. It is shown that there exists a spanning tree of
degree four whose weight is at most 1.25 times the weight of a minimum
spanning tree. These results solve open problems posed by Papadimitriou
and Vazirani.  Moreover, if a minimum spanning tree is given as part
of the input, the trees can be computed in $O(n)$ time.

The results are generalized to points in higher dimensions. It is
shown that for any $d \ge 3$, an arbitrary collection of points in
$\DimD$ contains a spanning tree of degree three, whose weight is at
most 5/3 times the weight of a minimum spanning tree.            This
is the first paper that  achieves factors better than two for these
problems.

\end{abstract}

AMS CLASSIFICATION:
05C05, 05C10, 05C85, 65Y25, 68Q20, 68R10, 68U05, 90C27, 90C35.


\section{Introduction}
Given $n$ points in the plane, how do we find a 
spanning tree of minimum weight among those 
in which each vertex has degree at most $K$?
Here the weight of an edge between two points is defined to
be the Euclidean distance between them.
This problem is referred to as the {\bf Euclidean degree-$K$ spanning tree
problem} and is a generalization of the Hamilton Path problem
which is known to be NP-hard~\cite{GJ,IPS}.
When $K=3$, it was shown to be NP-hard by
Papadimitriou and Vazirani~\cite{PV}, who conjectured that it is
NP-hard for $K=4$ as well. When $K=5$, the problem can be solved in
polynomial time~\cite{MS}.

This paper addresses the problem of computing low weight degree-$K$
spanning trees for $K>2$. In any metric space, it is known
that there always exists a spanning tree of degree $2$ whose cost is
at most twice the cost of a minimum spanning tree (MST). This is shown
by taking an Euler tour of an MST (in which each edge is taken twice)
and producing a Hamilton tour by short-cutting the Euler tour. In the
case of general metric spaces, it is easy to generate examples in
which the ratio of a shortest Hamilton path to the weight of a minimum
spanning tree is arbitrarily close to two.
But such examples do not
translate to points in $\DimD$. In view of this, Papadimitriou and
Vazirani~\cite{PV} posed the problem of obtaining factors better than
two for the Euclidean degree-$K$ spanning tree problem.  It should be
noted that in the special case of $K=2$, Christofides~\cite{Chr} gave
a simple and elegant polynomial time approximation algorithm with an
approximation ratio of 1.5 for computing a traveling salesperson tour
for points satisfying the triangle inequality (points in a metric space).

\subsection{Our Contributions}

In this paper, we show that for an arbitrary collection of $n$ points
in the plane, there exists a degree-3 spanning tree whose weight is at
most 1.5 times the weight of a minimum spanning tree. We also show
that there exists a degree-4 spanning tree whose weight is at most
1.25 times the weight of a minimum spanning tree.  This solves an
open problem posed by Papadimitriou and Vazirani~\cite{PV}.

Moreover, if a minimum spanning tree is given as part of the input,
the trees can be computed in $O(n)$ time.  Note that our bound of 1.5
for the degree-3 spanning tree problem is an ``absolute'' guarantee
(based on the weight of an MST) as opposed to a ``relative'' guarantee
for the degree-2 spanning tree obtained by Christofides~\cite{Chr}
(based on the weight of an optimal solution).

We also generalize our results to points in higher
dimensions. We show that for any $d \ge 2$, an arbitrary collection of
points in $\DimD$ contains a degree-3 spanning tree whose weight is at
most 5/3 times the weight of a minimum spanning tree.            This
is the first paper that  achieves factors better than two for these
problems.

\subsection{Significance of Our Results}

Many approximation algorithms      make use of the
triangle inequality to obtain approximate solutions to NP-hard problems.
These algorithms typically involve a ``short-cutting'' step where
the triangle inequality is used to bound the cost of the obtained solution. 
Examples include 
Christofides' heuristic for the traveling salesperson problem~\cite{Chr},
biconnectivity augmentation~\cite{FJ},
approximate weighted matching~\cite{GW},
prize-collecting traveling salesperson~\cite{BGSW},
and bounded-degree subgraphs which have 
low weight and small bottleneck cost~\cite{RMRRH}.

A question of general interest is how to obtain improved
approximation algorithms for such problems when the points 
come from a Euclidean, as opposed to arbitrary, metric space.
This requires making use of more than just the triangle inequality.
Surprisingly, for most problems, improved algorithms are not known.
(A notable exception is the famous Euclidean Steiner tree problem
\cite{DuH,DZF}.)
We use rudimentary geometric techniques to obtain an improved algorithm 
for the Euclidean degree-$K$ spanning tree problem.

The key to our method is to give short-cutting steps that
are provably better than implied by the triangle inequality alone.
Lemma~\ref{triangle-lemma}, which bounds the perimeter of an arbitrary
triangle in terms of distances to its vertices from any point,
is typical of the techniques that we use to get better bounds.

\subsection{Related Work}

Papadimitriou and Vazirani showed that any MST whose vertices have integer
co-ordinates has maximum degree at most five \cite{PV}.
Monma and Suri~\cite{MS} showed that for {\em every} set of points 
in the plane, there exists a degree-5 MST.

Many recent works have given algorithms to find subgraphs of bounded
degree that simultaneously satisfy other given constraints.
A polynomial-time algorithm to find a spanning tree or a Steiner tree
of a given subset of vertices in a graph, with degree at most one more
than minimum was given by F\"urer and Raghavachari~\cite{FR}.
This was extended to weighted graphs by Fischer \cite{Fi}.
He showed how to find minimum spanning trees whose degree is within a
constant multiplicative factor plus an additive $O(\log n)$ of the
optimal degree. The degree bound is improved further in the case when
the number of different edge weights is bounded by a constant.
Ravi, Marathe, Ravi, Rosenkrantz and Hunt~\cite{RMRRH} consider
the problem of computing bounded-degree subgraphs satisfying given
connectivity properties in a graph  whose edge weights satisfy the triangle
inequality. They give efficient algorithms for computing subgraphs
which have low weight and small bottleneck cost.
Salowe \cite{Sal}, and Das and
Heffernan \cite{DH} consider the problem of computing bounded-degree
graph spanners 
and provide algorithms for computing them.
Robins and Salowe \cite{RoS} study the maximum degrees of minimum 
spanning trees under various metrics.

\section{Preliminaries}

Let $V = \{v_1, \ldots, v_n\}$ be a set of $n$ points in the plane.
Let $G$ be the complete graph induced by $V$,
where the weight of an edge is the Euclidean distance between its
endpoints. We use the terms points and vertices interchangeably. Let
$\Edge{uv}$ be the Euclidean distance between vertices $u$ and $v$.
Let $T_{\min}$ be a            minimum spanning tree (MST) of the
points in $V$. Let $w(T)$ denote the total weight of a spanning tree
$T$.  Let $T_k$ denote a spanning tree in which every vertex has
degree at most $k$.
Let deg$_T(v)$ be the degree of a vertex $v$ in the tree $T$.
Let $\Delta ABC$ denote the triangle formed by points $A,B$ and $C$.
Let $\Angle{ABC}$ denote the angle formed at $B$ between
line segments $AB$ and $BC$.
Let $\Perimeter{ABC}$ denote the perimeter of $\Delta ABC$;
and more generally, let $\Perimeter{v_1v_2\ldots v_k}$ denote the
perimeter of the polygon formed by the line segments $v_iv_{i+1}$ for
$1\le i\le k$, where $v_{k+1}=v_1$.

In this paper we prove the following:
for an arbitrary set of points in $\Dim{2}$,
\begin{eqnarray}
 \exists T_3: \ \ w(T_3) &\leq& 1.5 \times w(T_{\min}) \label{deg3-result} \\
 \exists T_4: \ \ w(T_4) &\leq& 1.25 \times w(T_{\min}) \label{deg4-result}
\end{eqnarray}

For an arbitrary set of points in $\DimD$ ($d>2$),
 \begin{eqnarray}
 \exists T_3: \ \ w(T_3) &\leq& \frac{5}{3} \times 
                w(T_{\min}) \label{DimD-result}
 \end{eqnarray}

\section{Points in the plane}

We first consider the case of $\Dim{2}$ -- points in the plane.
We first note  some useful properties of 
minimum spanning trees in $\DimD$.

\begin{proposition}[\cite{PV}] \label{mst-angle-lemma}
Let $AB$ and $BC$ be two edges incident to a point $B$ in a minimum
spanning tree of a set of points in $\DimD$. Then $\Angle{ABC}$ is a
largest angle in $\Delta ABC$.
\end{proposition}

\begin{corollary} \label{sixty-degree-theorem}
                  \label{convexity-corollary}
Let $AB$ and $BC$ be two edges incident to a point $B$ in a minimum
spanning tree of a set of points in $\DimD$. 
Then
\begin{itemize}
\item $\Angle{ABC} \ge 60^\circ$
\item $\Angle{BAC}, \Angle{BCA} \le 90^\circ$.
\end{itemize}
\end{corollary}

\subsection{An upper bound on the perimeter of a triangle}

We now prove an upper bound on the perimeter of an arbitrary triangle
in terms of distances to its vertices from an arbitrary point.
This lemma is useful in proving the performances of our algorithms.
The lemma is also interesting in its own right and we believe
that it and the associated techniques will be useful in other
geometrical problems.

\begin{lemma}
\label{triangle-lemma}
   Let $X$, $A$, $B$, and $C$ be points in $\DimD$
   with $\Edge{XA} \leq \Edge{XB},     \Edge{XC}$. Then
\begin{equation}
 \Perimeter{ABC}  \leq (3\sqrt{3}-4) \Edge{XA}
        + 2 (\Edge{XB} + \Edge{XC}). \label{perimeter-eq}
\end{equation}
\end{lemma}
  Note that $3\sqrt{3}-4 \approx 1.2$.
  Recall that $\Perimeter{ABC}$ is the perimeter of the triangle
  and $\Edge{XY}$ is the distance from $X$ to $Y$.
  
\begin{figure}[htbp]
 \begin{center}
\setlength{\unitlength}{0.01250000in}
\begingroup\makeatletter\ifx\SetFigFont\undefined
\def\x#1#2#3#4#5#6#7\relax{\def\x{#1#2#3#4#5#6}}%
\expandafter\x\fmtname xxxxxx\relax \def\y{splain}%
\ifx\x\y   
\gdef\SetFigFont#1#2#3{%
  \ifnum #1<17\tiny\else \ifnum #1<20\small\else
  \ifnum #1<24\normalsize\else \ifnum #1<29\large\else
  \ifnum #1<34\Large\else \ifnum #1<41\LARGE\else
     \huge\fi\fi\fi\fi\fi\fi
  \csname #3\endcsname}%
\else
\gdef\SetFigFont#1#2#3{\begingroup
  \count@#1\relax \ifnum 25<\count@\count@25\fi
  \def\x{\endgroup\@setsize\SetFigFont{#2pt}}%
  \expandafter\x
    \csname \romannumeral\the\count@ pt\expandafter\endcsname
    \csname @\romannumeral\the\count@ pt\endcsname
  \csname #3\endcsname}%
\fi
\fi\endgroup
\begin{picture}(216,162)(0,-10)
\put(73,131){\blacken\ellipse{4}{4}}
\put(73,131){\ellipse{4}{4}}
\put(19,12){\blacken\ellipse{4}{4}}
\put(19,12){\ellipse{4}{4}}
\put(186,9){\blacken\ellipse{4}{4}}
\put(186,9){\ellipse{4}{4}}
\put(80,84){\blacken\ellipse{4}{4}}
\put(80,84){\ellipse{4}{4}}
\put(49,47){\blacken\ellipse{4}{4}}
\put(49,47){\ellipse{4}{4}}
\put(121,57){\blacken\ellipse{4}{4}}
\put(121,57){\ellipse{4}{4}}
\path(73,131)(186,9)
\path(73,131)(186,9)
\path(186,9)(19,12)
\path(186,9)(19,12)
\path(19,12)(73,131)
\path(19,12)(73,131)
\dashline{4.000}(74,130)(80,85)
\dashline{4.000}(81,84)(19,12)
\dashline{4.000}(81,84)(187,9)
\dottedline{3}(73,131)(49,47)
\dottedline{3}(49,46)(121,56)(74,131)
\put(0,5){\makebox(0,0)[lb]{\smash{{{\SetFigFont{10}{12.0}{rm}$B$}}}}}
\put(199,2){\makebox(0,0)[lb]{\smash{{{\SetFigFont{10}{12.0}{rm}$C$}}}}}
\put(47,31){\makebox(0,0)[lb]{\smash{{{\SetFigFont{10}{12.0}{rm}$B'$}}}}}
\put(82,91){\makebox(0,0)[lb]{\smash{{{\SetFigFont{10}{12.0}{rm}$X$}}}}}
\put(77,136){\makebox(0,0)[lb]{\smash{{{\SetFigFont{10}{12.0}{rm}$A$}}}}}
\put(112,41){\makebox(0,0)[lb]{\smash{{{\SetFigFont{10}{12.0}{rm}$C'$}}}}}
\end{picture}

 \end{center}
 \caption{Shrinking to obtain canonical form.}
 \label{triangle}
 \end{figure}
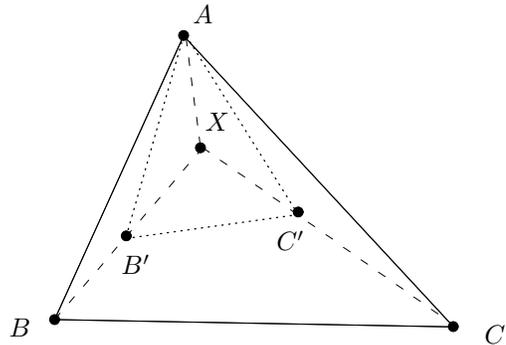

\begin{proof}
Let $B'$ and $C'$ be points on $XB$ and $XC$ respectively such that
$\Edge{XA}=\Edge{XB'}=\Edge{XC'}$ (see Fig.~\ref{triangle}).
First we observe that the lemma is true if it is true for the points
$X, A, B'$ and $C'$. This follows because by the triangle inequality,
\[ \Perimeter{ABC} \le \Perimeter{AB'C'} + 2\Edge{BB'} + 2\Edge{CC'}. \]
By our assumption,
\[ \Perimeter{AB'C'} \le (3\sqrt{3}-4) \Edge{XA}
        + 2 (\Edge{XB'} + \Edge{XC'}). \]
Combining the two inequalities yields the desired result.
Therefore in the rest of the proof, we show that the lemma is true
when the ``arms'' $\Edge{XA}$, $\Edge{XB'}$ and $\Edge{XC'}$ are equal.

It is not very difficult to see that to maximize the perimeter of the
triangle, $X$ will be in the plane defined by $A, B'$ and $C'$, and thus
$X$ is at the center of a circle passing through $A, B'$ and $C'$.

By scaling, it suffices to consider the case when the circle has unit radius.
In this case, the right-hand side of (\ref{perimeter-eq}) 
is exactly $3\sqrt{3}$.  
Thus, it suffices to show that the maximum perimeter 
achieved by any triangle whose vertices lie on a unit circle 
is $3\sqrt{3}$.
This is easily proved \cite{Li}.
\end{proof}

Note that in an arbitrary metric space
it is possible to have an (equilateral) triangle of perimeter six
and a point $X$ at distance one from each vertex.

\subsection{Spanning trees of degree three}

We now assume that we are given a Euclidean minimum spanning tree $T$
of degree at most five. We show how to convert $T$ into a tree of
degree at most three. The weight of the resulting tree is at most 1.5
times the weight of $T$.

\noindent
{\bf High Level Description:} 
The tree $T$ is rooted at an arbitrary leaf vertex.
Since $T$ is a degree-5 tree, once it is rooted at a leaf,
each vertex has at most four children.
For each vertex $v$, the shortest path $P_v$ starting at $v$
and visiting every child of $v$ is computed.
The final tree $T_3$ consists of the union of the paths $\{P_v\}$.
Fig.~\ref{tree3-alg} gives the above algorithm.
In analyzing the algorithm, we think of each vertex $v$
as replacing its edges from its children with the path $P_v$.
The above technique of ``shortcutting'' the children of a vertex by
``stringing'' them together has been known before, especially in the
context of computing degree-3 trees in metric spaces (see \cite{RMRRH,Sal}).

 \begin{figure}[ht]
 \begin{tabAlgorithm}{{\sc Tree-3}$(V,T)$}
 {Find a degree 3 tree of  $V$.}
 \algline Root the MST $T$ at a leaf vertex $r$.
 \algline {\bf For} \= each vertex $v \in V$ do
 \algline       \> Compute $P_v$, the shortest path starting 
        at $v$ and visiting all the children of $v$.
 \algline Return $T_3$, the tree formed by the union of the paths $\{P_v\}$.
 \end{tabAlgorithm}
 \caption{Algorithm to find a degree 3 tree.}
 \label{tree3-alg}
 \end{figure}

\noindent{\bf Note:}
Typically, the initial MST has very few nodes with degree greater than
three~\cite{bentley}. 
In practice, it is worth modifying the algorithm 
to scan the vertices in preorder, maintaining the partial tree $T_3$ of
edges added so far, and to add paths to $T_3$ as follows.
When considering a vertex $v$, 
if the degree of $v$ in the partial $T_3$ is two,
add the path $P_v$ as described in the algorithm.
Otherwise its degree is one and, in this case, relax the requirement
that the added path must start at $v$. 
That is, add the shortest path that visits $v$ and all of $v$'s
children to $T_3$ (see \S\ref{sec-degree-four}).
This modification will never increase the cost of the resulting tree,
but may offer substantially lighter trees in practice.

\begin{lemma}
The algorithm in Fig.~\ref{tree3-alg} outputs a spanning tree of
degree three.
\end{lemma}

\begin{proof}
An easy proof by induction shows that the union of the paths forms a tree.
Each vertex $v$ is on at most two paths and is an interior vertex
of at most one path.
\end{proof}

\begin{lemma} \label{deg3-theorem}
Let $v$ be a vertex in an MST $T$ of a set of points in $\Dim{2}$.
Let $P_v$ be a shortest path visiting $\{v\}\ \cup$ child$_T(v)$
with $v$ as one of its endpoints.
\[ w(P_v) \leq 1.5 \times \sum_{v_i \in \mbox{child}_T(v)} \Edge{vv_i}.\]
\end{lemma}

By the above lemma, each path $P_v$ has weight at most $1.5$ times
the weight of the edges it replaces.  Thus,
\begin{theorem} \label{deg3-R2-theorem}
Let $T$ be a minimum spanning tree of a set of points in $\Dim{2}$.
Let $T_3$ be the spanning tree output by the algorithm in Fig.~\ref{tree3-alg}.
\[ w(T_3) \leq 1.5 \times w(T).\]
\end{theorem}

\noindent{\it Proof of Lemma~\ref{deg3-theorem}. }
We consider the various cases that arise depending
on the number of children of $v$.
The cases when $v$ has no children or exactly one child are trivial.

\noindent
{\em Case 1: $v$ has 2 children, $v_1,v_2$. }
There are two possible paths for $P_v$, namely $P_1=[v,v_1,v_2]$ and 
$P_2=[v,v_2,v_1]$. 
Clearly,
\[
 w(P_v) = \min(w(P_1), w(P_2))
        \leq \frac{w(P_1) + w (P_2)}{2}
        =    \frac{\Edge{vv_1}}{2} + \frac{\Edge{vv_2}}{2} + \Edge{v_1v_2}
        \leq 1.5 \ (\Edge{vv_1} + \Edge{vv_2}). 
\]

\begin{figure}[htbp]
 \begin{center}
\setlength{\unitlength}{0.0085in}
\begingroup\makeatletter\ifx\SetFigFont\undefined
\def\x#1#2#3#4#5#6#7\relax{\def\x{#1#2#3#4#5#6}}%
\expandafter\x\fmtname xxxxxx\relax \def\y{splain}%
\ifx\x\y   
\gdef\SetFigFont#1#2#3{%
  \ifnum #1<17\tiny\else \ifnum #1<20\small\else
  \ifnum #1<24\normalsize\else \ifnum #1<29\large\else
  \ifnum #1<34\Large\else \ifnum #1<41\LARGE\else
     \huge\fi\fi\fi\fi\fi\fi
  \csname #3\endcsname}%
\else
\gdef\SetFigFont#1#2#3{\begingroup
  \count@#1\relax \ifnum 25<\count@\count@25\fi
  \def\x{\endgroup\@setsize\SetFigFont{#2pt}}%
  \expandafter\x
    \csname \romannumeral\the\count@ pt\expandafter\endcsname
    \csname @\romannumeral\the\count@ pt\endcsname
  \csname #3\endcsname}%
\fi
\fi\endgroup
\begin{picture}(630,215)(0,-10)
\put(125,38){\blacken\ellipse{4}{4}}
\put(125,38){\ellipse{4}{4}}
\put(5,38){\blacken\ellipse{4}{4}}
\put(5,38){\ellipse{4}{4}}
\put(65,98){\blacken\ellipse{4}{4}}
\put(65,98){\ellipse{4}{4}}
\put(165,38){\blacken\ellipse{4}{4}}
\put(165,38){\ellipse{4}{4}}
\put(225,98){\blacken\ellipse{4}{4}}
\put(225,98){\ellipse{4}{4}}
\put(65,178){\blacken\ellipse{4}{4}}
\put(65,178){\ellipse{4}{4}}
\put(285,38){\blacken\ellipse{4}{4}}
\put(285,38){\ellipse{4}{4}}
\put(225,178){\blacken\ellipse{4}{4}}
\put(225,178){\ellipse{4}{4}}
\put(325,38){\blacken\ellipse{4}{4}}
\put(325,38){\ellipse{4}{4}}
\put(385,178){\blacken\ellipse{4}{4}}
\put(385,178){\ellipse{4}{4}}
\put(385,98){\blacken\ellipse{4}{4}}
\put(385,98){\ellipse{4}{4}}
\put(445,38){\blacken\ellipse{4}{4}}
\put(445,38){\ellipse{4}{4}}
\put(485,38){\blacken\ellipse{4}{4}}
\put(485,38){\ellipse{4}{4}}
\put(545,98){\blacken\ellipse{4}{4}}
\put(545,98){\ellipse{4}{4}}
\put(545,178){\blacken\ellipse{4}{4}}
\put(545,178){\ellipse{4}{4}}
\put(605,38){\blacken\ellipse{4}{4}}
\put(605,38){\ellipse{4}{4}}
\path(65,98)(5,38)(125,38)(65,178)
\path(605,38)(485,38)(545,178)(545,98)
\path(223,99)(165,38)(225,178)(285,38)
\path(385,98)(445,38)(325,38)(385,178)
\put(0,18){\makebox(0,0)[lb]{\smash{{{\SetFigFont{12}{14.4}{rm}$v_1$}}}}}
\put(55,3){\makebox(0,0)[lb]{\smash{{{\SetFigFont{12}{14.4}{rm}$P_1$}}}}}
\put(65,108){\makebox(0,0)[lb]{\smash{{{\SetFigFont{12}{14.4}{rm}$v$}}}}}
\put(60,188){\makebox(0,0)[lb]{\smash{{{\SetFigFont{12}{14.4}{rm}$v_3$}}}}}
\put(160,18){\makebox(0,0)[lb]{\smash{{{\SetFigFont{12}{14.4}{rm}$v_1$}}}}}
\put(120,18){\makebox(0,0)[lb]{\smash{{{\SetFigFont{12}{14.4}{rm}$v_2$}}}}}
\put(215,3){\makebox(0,0)[lb]{\smash{{{\SetFigFont{12}{14.4}{rm}$P_2$}}}}}
\put(220,108){\makebox(0,0)[lb]{\smash{{{\SetFigFont{12}{14.4}{rm}$v$}}}}}
\put(215,188){\makebox(0,0)[lb]{\smash{{{\SetFigFont{12}{14.4}{rm}$v_3$}}}}}
\put(280,18){\makebox(0,0)[lb]{\smash{{{\SetFigFont{12}{14.4}{rm}$v_2$}}}}}
\put(320,18){\makebox(0,0)[lb]{\smash{{{\SetFigFont{12}{14.4}{rm}$v_1$}}}}}
\put(375,3){\makebox(0,0)[lb]{\smash{{{\SetFigFont{12}{14.4}{rm}$P_3$}}}}}
\put(375,188){\makebox(0,0)[lb]{\smash{{{\SetFigFont{12}{14.4}{rm}$v_3$}}}}}
\put(380,108){\makebox(0,0)[lb]{\smash{{{\SetFigFont{12}{14.4}{rm}$v$}}}}}
\put(440,18){\makebox(0,0)[lb]{\smash{{{\SetFigFont{12}{14.4}{rm}$v_2$}}}}}
\put(480,18){\makebox(0,0)[lb]{\smash{{{\SetFigFont{12}{14.4}{rm}$v_1$}}}}}
\put(535,3){\makebox(0,0)[lb]{\smash{{{\SetFigFont{12}{14.4}{rm}$P_4$}}}}}
\put(600,18){\makebox(0,0)[lb]{\smash{{{\SetFigFont{12}{14.4}{rm}$v_2$}}}}}
\put(535,188){\makebox(0,0)[lb]{\smash{{{\SetFigFont{12}{14.4}{rm}$v_3$}}}}}
\put(540,78){\makebox(0,0)[lb]{\smash{{{\SetFigFont{12}{14.4}{rm}$v$}}}}}
\end{picture}

 \end{center}
 \caption{$T_3$, three children}
 \label{d3-3c}
 \end{figure}
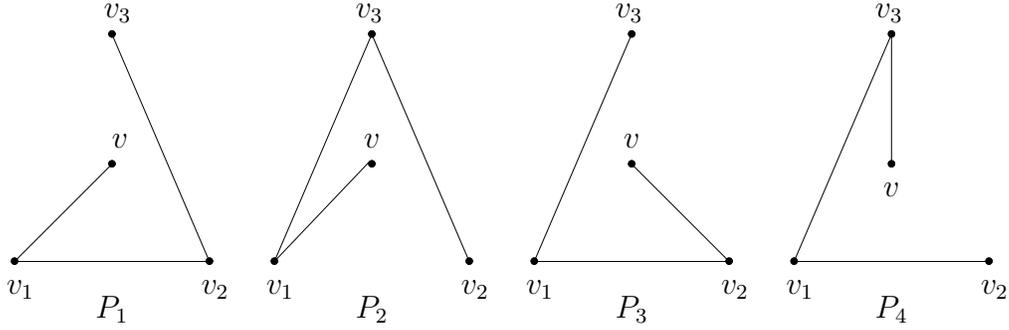

\noindent
{\em Case 2: $v$ has 3 children, $v_1,v_2,v_3$. }
Let $v_1$ be the child that is nearest to $v$.
Consider the following four paths (see Fig.~\ref{d3-3c}):
$P_1 = [v, v_1, v_2, v_3], P_2 = [v, v_1, v_3, v_2], 
        P_3 = [v, v_2, v_1, v_3]$ and $P_4 = [v, v_3, v_1, v_2]$.  

The path $P_v$
is at most as heavy as the lightest of
$\{P_1,P_2,P_3,P_4\}$. The weight of the lightest of these paths
is at most any convex combination of the weights of the paths.
Specifically,
\[
 w(P_v) \le \min(w(P_1), w(P_2), w(P_3), w(P_4))
        \leq \frac{w(P_1)}{3} 
        + \frac{w(P_2)}{3} + \frac{w(P_3)}{6} + \frac{w(P_4)}{6}.
\]

We will now prove that
\[ \frac{w(P_1)}{3} + \frac{w(P_2)}{3} + \frac{w(P_3)}{6} + \frac{w(P_4)}{6}
        \leq 1.5 \ (\Edge{vv_1} + \Edge{vv_2} + \Edge{vv_3}).\]

   This simplifies to
\[ \Edge{v_1v_2} + \Edge{v_2v_3} + \Edge{v_3v_1} \leq
   1.25 \ \Edge{vv_1} + 2( \Edge{vv_2} + \Edge{vv_3}), \]
   which follows from Lemma~\ref{triangle-lemma}.

\noindent
{\em Case 3: $v$ has 4 children, $v_1, v_2, v_3, v_4$, 
ordered clockwise around $v$. }
Let $v'$ be the point of intersection of the 
diagonals $\Edge{v_1 v_3}$ and $\Edge{v_2 v_4}$.
Note that the diagonals do intersect because
the polygon $v_1v_2v_3v_4$ is convex
(follows from 
Corollary~\ref{convexity-corollary}).

\begin{figure}[htbp]
 \begin{center}
\setlength{\unitlength}{0.01250000in}
\begingroup\makeatletter\ifx\SetFigFont\undefined
\def\x#1#2#3#4#5#6#7\relax{\def\x{#1#2#3#4#5#6}}%
\expandafter\x\fmtname xxxxxx\relax \def\y{splain}%
\ifx\x\y   
\gdef\SetFigFont#1#2#3{%
  \ifnum #1<17\tiny\else \ifnum #1<20\small\else
  \ifnum #1<24\normalsize\else \ifnum #1<29\large\else
  \ifnum #1<34\Large\else \ifnum #1<41\LARGE\else
     \huge\fi\fi\fi\fi\fi\fi
  \csname #3\endcsname}%
\else
\gdef\SetFigFont#1#2#3{\begingroup
  \count@#1\relax \ifnum 25<\count@\count@25\fi
  \def\x{\endgroup\@setsize\SetFigFont{#2pt}}%
  \expandafter\x
    \csname \romannumeral\the\count@ pt\expandafter\endcsname
    \csname @\romannumeral\the\count@ pt\endcsname
  \csname #3\endcsname}%
\fi
\fi\endgroup
\begin{picture}(400,210)(0,-10)
\put(360,113){\blacken\ellipse{4}{4}}
\put(360,113){\ellipse{4}{4}}
\put(300,173){\blacken\ellipse{4}{4}}
\put(300,173){\ellipse{4}{4}}
\put(280,53){\blacken\ellipse{4}{4}}
\put(280,53){\ellipse{4}{4}}
\put(300,133){\blacken\ellipse{4}{4}}
\put(300,133){\ellipse{4}{4}}
\put(220,133){\blacken\ellipse{4}{4}}
\put(220,133){\ellipse{4}{4}}
\put(80,53){\blacken\ellipse{4}{4}}
\put(80,53){\ellipse{4}{4}}
\put(20,133){\blacken\ellipse{4}{4}}
\put(20,133){\ellipse{4}{4}}
\put(160,113){\blacken\ellipse{4}{4}}
\put(160,113){\ellipse{4}{4}}
\put(100,173){\blacken\ellipse{4}{4}}
\put(100,173){\ellipse{4}{4}}
\put(100,133){\blacken\ellipse{4}{4}}
\put(100,133){\ellipse{4}{4}}
\path(100,133)(20,133)(100,173)
        (160,113)(80,53)
\path(300,133)(360,113)(300,173)
        (220,133)(280,53)
\put(80,3){\makebox(0,0)[lb]{\smash{{{\SetFigFont{12}{14.4}{rm}$P_1$}}}}}
\put(75,38){\makebox(0,0)[lb]{\smash{{{\SetFigFont{12}{14.4}{rm}$v_3$}}}}}
\put(90,183){\makebox(0,0)[lb]{\smash{{{\SetFigFont{12}{14.4}{rm}$v_1$}}}}}
\put(290,183){\makebox(0,0)[lb]{\smash{{{\SetFigFont{12}{14.4}{rm}$v_1$}}}}}
\put(275,38){\makebox(0,0)[lb]{\smash{{{\SetFigFont{12}{14.4}{rm}$v_3$}}}}}
\put(280,3){\makebox(0,0)[lb]{\smash{{{\SetFigFont{12}{14.4}{rm}$P_2$}}}}}
\put(370,108){\makebox(0,0)[lb]{\smash{{{\SetFigFont{12}{14.4}{rm}$v_2$}}}}}
\put(170,103){\makebox(0,0)[lb]{\smash{{{\SetFigFont{12}{14.4}{rm}$v_2$}}}}}
\put(95,118){\makebox(0,0)[lb]{\smash{{{\SetFigFont{12}{14.4}{rm}$v$}}}}}
\put(295,118){\makebox(0,0)[lb]{\smash{{{\SetFigFont{12}{14.4}{rm}$v$}}}}}
\put(0,128){\makebox(0,0)[lb]{\smash{{{\SetFigFont{12}{14.4}{rm}$v_4$}}}}}
\put(200,128){\makebox(0,0)[lb]{\smash{{{\SetFigFont{12}{14.4}{rm}$v_4$}}}}}
\end{picture}

 \end{center}
 \caption{$T_3$, four children}
 \label{d3-4c}
 \end{figure}
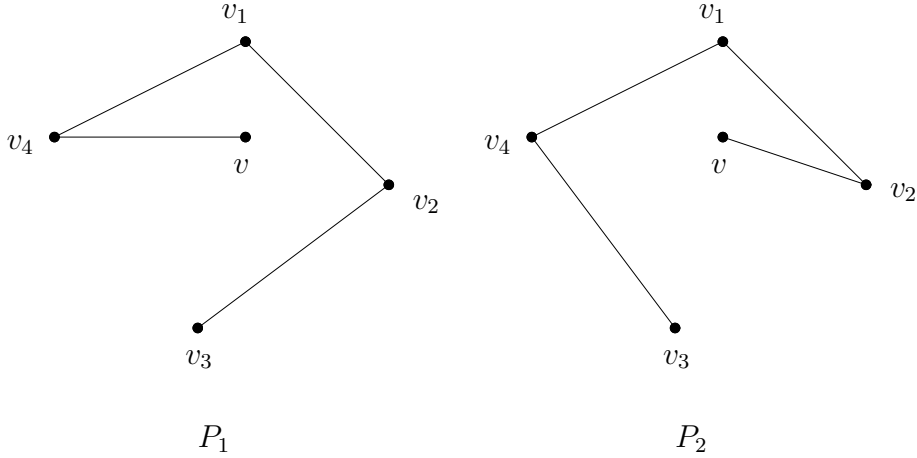

Let $v_3$ be the point that is furthest from $v'$, among
$\{v_1,v_2,v_3,v_4\}$. 
Consider the following two paths (see Fig.~\ref{d3-4c}):
$P_1 = [v, v_4, v_1, v_2, v_3], P_2 = [v, v_2, v_1, v_4, v_3]$.

   Clearly,
\[ w(P_v) \le \min(w(P_1), w(P_2)) \leq \frac{w(P_1)}{2} + \frac{w(P_2)}{2}.\]

We will show that 
\[  \frac{1}{2} (w(P_1) +  w(P_2)) \leq 
        1.5 (\Edge{vv_1} + \Edge{vv_2} + \Edge{vv_3} + \Edge{vv_4}).
\]

This simplifies to
\begin{equation} \label{quad-equation}
   \Perimeter{v_1v_2v_3v_4} + (\Edge{v_1v_2} + \Edge{v_1v_4})
        \leq 3( \Edge{vv_1} + \Edge{vv_3}) + 2(\Edge{vv_2} + \Edge{vv_4}).
\end{equation}

We will first prove that
\begin{equation} \label{quad-equation2}
  \Perimeter{v_1v_2v_3v_4} + (\Edge{v_1v_2} + \Edge{v_1v_4})
        \leq 3( \Edge{v'v_1} + \Edge{v'v_3}) + 2(\Edge{v'v_2} + \Edge{v'v_4}).
\end{equation}

Once we prove (\ref{quad-equation2}), by the triangle inequality we can
conclude that (\ref{quad-equation}) is true.
(Since $\Edge{vv_1} + \Edge{vv_3} \geq \Edge{v_1v_3} = 
\Edge{v'v_1} + \Edge{v'v_3}$ and 
$\Edge{vv_2} + \Edge{vv_4} \geq \Edge{v_2v_4} = \Edge{v'v_2} + \Edge{v'v_4}$.)

We prove (\ref{quad-equation2}) by contradiction. Suppose there exists a set of
points which does not satisfy (\ref{quad-equation2}).  Suppose we
shrink $v'v_3$ by $\delta$. The left side of the above inequality
decreases by at most $2\delta$, whereas the right side of the
inequality decreases by exactly $3\delta$. Therefore as we shrink
$v'v_3$, the inequality stays violated. Suppose $v'v_3$ shrinks and
becomes equal to another edge $v'v_i$ for some $i\in\{1,2,4\}$. We now
shrink both $v'v_3$ and $v'v_i$ simultaneously at the same rate. Again
it is easy to show that the inequality continues to be violated as
$v'v_3$ and $v'v_i$ shrink. Hence we reach a configuration
where three of the edges are equal.  

Without loss of generality, the
length of the three edges is 1 and the length of the fourth edge is some
$\epsilon\le 1$. 

There are two cases to consider. The first is when $v'v_1 = \epsilon$
and the second is when $v'v_2 = \epsilon$. (The case when $v'v_4 =
\epsilon$ is the same as the second case.)

\begin{enumerate}
\item[{\it Case 3a.}] $v'v_1 = \epsilon$.
We wish to prove that 
\[ \Perimeter{v_1v_2v_3v_4} + (\Edge{v_1v_2} + \Edge{v_1v_4}) \leq 
7+3 \epsilon.\]
We want to show that the function $F(\epsilon) = 
\Perimeter{v_1v_2v_3v_4} + (\Edge{v_1v_2} + \Edge{v_1v_4}) - 7 - 3 \epsilon$
is non-positive in the domain $0 \leq \epsilon \leq 1$.
Simplifying, we get
\[ F(\epsilon) = 
2 \Edge{v_1v_2} + \Edge{v_2v_3} + \Edge{v_3v_4} +2 \Edge{v_1v_4}
 - 7 - 3 \epsilon.
\]

Each of $\Edge{v_iv_j}$ in the definition of $F$ is a convex function of
$\epsilon$ due to the following reason. Let $p$ be the point closest to
$v_j$ on the line connecting $v_i$ and $v'$. Observe that as $v_i$ moves
towards $v'$, $\Edge{v_iv_j}$ decreases if $v_i$ is moving towards $p$
and increases otherwise. Since $F$ is a sum of convex functions
minus a linear function, it is a convex function of $\epsilon$.
Therefore it is maximized at either $\epsilon=0$ or $\epsilon =1$.

When $\epsilon=1$, all four points are at the same distance from $v'$.
If angle $\Angle{v_4v'v_1} = \alpha$ then $F$ can be written
as a function of a single variable $\alpha$ and it can be verified that $F$ 
reaches a maximum value of $10\sqrt{0.8}-10$, which
is non-positive.

When $\epsilon=0$, $\Edge{v_1v_2}= \Edge{v_1v_4} = 1$. Simplifying
we get $F=\Edge{v_2v_3} + \Edge{v_3v_4} - 3$,
and it reaches a maximum value of $2\sqrt{2}-3$, which is
non-positive (when $\epsilon=0$, note that $v_1$ is the midpoint of the
line segment $v_2v_4$).

\item[{\it Case 3b.}] $v'v_2 = \epsilon$.
We wish to prove that 
\[ \Perimeter{v_1v_2v_3v_4} + (\Edge{v_1v_2} + \Edge{v_1v_4}) \leq 
8+2 \epsilon.\]
We want to show that the function $F'(\epsilon) = 
\Perimeter{v_1v_2v_3v_4} + (\Edge{v_1v_2} + \Edge{v_1v_4}) - 8 - 2 \epsilon$
is non-positive in the domain $0 \leq \epsilon \leq 1$.

As a function of $\epsilon$, function $F'$ is a sum of convex functions
minus a linear function, and thus is convex. Therefore it is maximized at
either $\epsilon=0$ or $\epsilon =1$.

The case $\epsilon=1$ leads to the same configuration as in Case 3a.

When $\epsilon=0$, $\Edge{v_1v_2}= \Edge{v_2v_3} = 1$.
Here $F'=2\Edge{v_1v_4} + \Edge{v_3v_4} - 5$.
If angle $\Angle{v_4v'v_1} = \alpha$, then $F'$ can be written
as a function of a single variable $\alpha$ and it can be verified that
$F'$ reaches a maximum value of $5\sqrt{0.8}-5$, which is non-positive.
\end{enumerate}

This concludes the proof of  Lemma~\ref{deg3-theorem}.
\myendproof

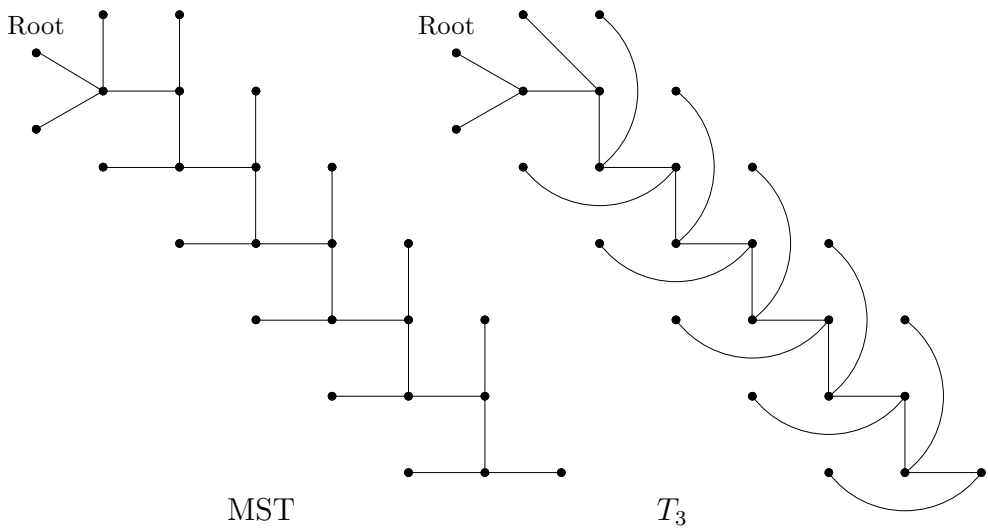
\begin{figure}[htbp]
 \begin{center}
\setlength{\unitlength}{0.01in}
\begingroup\makeatletter\ifx\SetFigFont\undefined
\def\x#1#2#3#4#5#6#7\relax{\def\x{#1#2#3#4#5#6}}%
\expandafter\x\fmtname xxxxxx\relax \def\y{splain}%
\ifx\x\y   
\gdef\SetFigFont#1#2#3{%
  \ifnum #1<17\tiny\else \ifnum #1<20\small\else
  \ifnum #1<24\normalsize\else \ifnum #1<29\large\else
  \ifnum #1<34\Large\else \ifnum #1<41\LARGE\else
     \huge\fi\fi\fi\fi\fi\fi
  \csname #3\endcsname}%
\else
\gdef\SetFigFont#1#2#3{\begingroup
  \count@#1\relax \ifnum 25<\count@\count@25\fi
  \def\x{\endgroup\@setsize\SetFigFont{#2pt}}%
  \expandafter\x
    \csname \romannumeral\the\count@ pt\expandafter\endcsname
    \csname @\romannumeral\the\count@ pt\endcsname
  \csname #3\endcsname}%
\fi
\fi\endgroup
\begin{picture}(513,285)(0,-10)
\put(280.000,228.000){\arc{100.000}{5.3559}{7.2105}}
\put(320.000,188.000){\arc{100.000}{5.3559}{7.2105}}
\put(360.000,148.000){\arc{100.000}{5.3559}{7.2105}}
\put(400.000,108.000){\arc{100.000}{5.3559}{7.2105}}
\put(440.000,68.000){\arc{100.000}{5.3559}{7.2105}}
\put(310.000,218.000){\arc{100.000}{0.6435}{2.4981}}
\put(350.000,178.000){\arc{100.000}{0.6435}{2.4981}}
\put(390.000,138.000){\arc{100.000}{0.6435}{2.4981}}
\put(430.000,98.000){\arc{100.000}{0.6435}{2.4981}}
\put(470.000,58.000){\arc{100.000}{0.6435}{2.4981}}
\put(210,28){\blacken\ellipse{4}{4}}
\put(210,28){\ellipse{4}{4}}
\put(250,28){\blacken\ellipse{4}{4}}
\put(250,28){\ellipse{4}{4}}
\put(290,28){\blacken\ellipse{4}{4}}
\put(290,28){\ellipse{4}{4}}
\put(250,68){\blacken\ellipse{4}{4}}
\put(250,68){\ellipse{4}{4}}
\put(250,108){\blacken\ellipse{4}{4}}
\put(250,108){\ellipse{4}{4}}
\put(210,68){\blacken\ellipse{4}{4}}
\put(210,68){\ellipse{4}{4}}
\put(170,68){\blacken\ellipse{4}{4}}
\put(170,68){\ellipse{4}{4}}
\put(210,108){\blacken\ellipse{4}{4}}
\put(210,108){\ellipse{4}{4}}
\put(210,148){\blacken\ellipse{4}{4}}
\put(210,148){\ellipse{4}{4}}
\put(170,108){\blacken\ellipse{4}{4}}
\put(170,108){\ellipse{4}{4}}
\put(130,108){\blacken\ellipse{4}{4}}
\put(130,108){\ellipse{4}{4}}
\put(170,148){\blacken\ellipse{4}{4}}
\put(170,148){\ellipse{4}{4}}
\put(170,188){\blacken\ellipse{4}{4}}
\put(170,188){\ellipse{4}{4}}
\put(130,148){\blacken\ellipse{4}{4}}
\put(130,148){\ellipse{4}{4}}
\put(90,148){\blacken\ellipse{4}{4}}
\put(90,148){\ellipse{4}{4}}
\put(130,188){\blacken\ellipse{4}{4}}
\put(130,188){\ellipse{4}{4}}
\put(130,228){\blacken\ellipse{4}{4}}
\put(130,228){\ellipse{4}{4}}
\put(90,188){\blacken\ellipse{4}{4}}
\put(90,188){\ellipse{4}{4}}
\put(50,188){\blacken\ellipse{4}{4}}
\put(50,188){\ellipse{4}{4}}
\put(90,228){\blacken\ellipse{4}{4}}
\put(90,228){\ellipse{4}{4}}
\put(90,268){\blacken\ellipse{4}{4}}
\put(90,268){\ellipse{4}{4}}
\put(50,228){\blacken\ellipse{4}{4}}
\put(50,228){\ellipse{4}{4}}
\put(50,268){\blacken\ellipse{4}{4}}
\put(50,268){\ellipse{4}{4}}
\put(15,248){\blacken\ellipse{4}{4}}
\put(15,248){\ellipse{4}{4}}
\put(15,208){\blacken\ellipse{4}{4}}
\put(15,208){\ellipse{4}{4}}
\put(430,28){\blacken\ellipse{4}{4}}
\put(430,28){\ellipse{4}{4}}
\put(470,28){\blacken\ellipse{4}{4}}
\put(470,28){\ellipse{4}{4}}
\put(510,28){\blacken\ellipse{4}{4}}
\put(510,28){\ellipse{4}{4}}
\put(470,68){\blacken\ellipse{4}{4}}
\put(470,68){\ellipse{4}{4}}
\put(470,108){\blacken\ellipse{4}{4}}
\put(470,108){\ellipse{4}{4}}
\put(430,68){\blacken\ellipse{4}{4}}
\put(430,68){\ellipse{4}{4}}
\put(390,68){\blacken\ellipse{4}{4}}
\put(390,68){\ellipse{4}{4}}
\put(430,108){\blacken\ellipse{4}{4}}
\put(430,108){\ellipse{4}{4}}
\put(430,148){\blacken\ellipse{4}{4}}
\put(430,148){\ellipse{4}{4}}
\put(390,108){\blacken\ellipse{4}{4}}
\put(390,108){\ellipse{4}{4}}
\put(350,108){\blacken\ellipse{4}{4}}
\put(350,108){\ellipse{4}{4}}
\put(390,148){\blacken\ellipse{4}{4}}
\put(390,148){\ellipse{4}{4}}
\put(390,188){\blacken\ellipse{4}{4}}
\put(390,188){\ellipse{4}{4}}
\put(350,148){\blacken\ellipse{4}{4}}
\put(350,148){\ellipse{4}{4}}
\put(310,148){\blacken\ellipse{4}{4}}
\put(310,148){\ellipse{4}{4}}
\put(350,188){\blacken\ellipse{4}{4}}
\put(350,188){\ellipse{4}{4}}
\put(350,228){\blacken\ellipse{4}{4}}
\put(350,228){\ellipse{4}{4}}
\put(310,188){\blacken\ellipse{4}{4}}
\put(310,188){\ellipse{4}{4}}
\put(270,188){\blacken\ellipse{4}{4}}
\put(270,188){\ellipse{4}{4}}
\put(310,228){\blacken\ellipse{4}{4}}
\put(310,228){\ellipse{4}{4}}
\put(310,268){\blacken\ellipse{4}{4}}
\put(310,268){\ellipse{4}{4}}
\put(270,228){\blacken\ellipse{4}{4}}
\put(270,228){\ellipse{4}{4}}
\put(270,268){\blacken\ellipse{4}{4}}
\put(270,268){\ellipse{4}{4}}
\put(235,248){\blacken\ellipse{4}{4}}
\put(235,248){\ellipse{4}{4}}
\put(235,208){\blacken\ellipse{4}{4}}
\put(235,208){\ellipse{4}{4}}
\path(210,148)(210,68)
\path(170,68)(250,68)
\path(250,108)(250,28)
\path(210,28)(290,28)
\path(210,108)(130,108)
\path(170,108)(170,188)
\path(170,148)(90,148)
\path(130,148)(130,228)
\path(130,188)(50,188)
\path(90,188)(90,268)
\path(90,228)(50,228)(50,268)
\path(50,228)(15,249)
\path(50,228)(15,208)
\path(270,228)(310,228)(310,188)
        (350,188)(350,148)(390,148)
        (390,108)(430,108)(430,68)
        (470,68)(470,28)(510,28)
\path(235,248)(270,228)
\path(235,208)(270,228)
\path(270,268)(310,228)
\put(115,3){\makebox(0,0)[lb]{\smash{{{\SetFigFont{12}{14.4}{rm}MST}}}}}
\put(340,3){\makebox(0,0)[lb]{\smash{{{\SetFigFont{12}{14.4}{rm}$T_3$}}}}}
\put(0,258){\makebox(0,0)[lb]{\smash{{{\SetFigFont{10}{12.0}{rm}Root}}}}}
\put(215,258){\makebox(0,0)[lb]{\smash{{{\SetFigFont{10}{12.0}{rm}Root}}}}}
\end{picture}

 \end{center}
 \caption{Bad example for algorithm in Fig.~\protect\ref{tree3-alg}.}
 \label{bad-ex}
 \end{figure}

The example in Fig.~\ref{bad-ex} shows that the 1.5 factor is tight
for the algorithm in Fig.~\ref{tree3-alg}, modified according to the
note following its description. The same example also shows that the
1.5 factor is tight for the unmodified algorithm since the unmodified
algorithm never outputs a lighter tree than the modified algorithm.
Each curved arc shown in Fig.~\ref{bad-ex} is actually a straight line,
and has been drawn curved for convenience. The vertex that is the child
of the root has three children, and is forced to drop one child. 
In doing so, the degree of its child goes to four, and it in turn
drops one of its children.
The algorithm could make choices in such a way that the changes
propagate through the tree and the tree $T_3$ output by the algorithm
may be as shown in the figure.
The ratio of the cost of the final solution to the cost
of the MST can be made arbitrarily close to 1.5.
See \S\ref{sec-conc} for a discussion on the worst case
ratio between degree-3 trees and minimum spanning trees.


\subsection{Spanning trees of degree four} \label{sec-degree-four}

We now assume that we are given a Euclidean minimum spanning tree in
which every vertex has degree at most 5. We show how to convert this
tree to a tree in which every vertex has degree at most~4.

\noindent
{\bf High Level Description:}
The basic idea is the same as in the previous algorithm.

 The difference is that we don't insist that each path $P_v$ start at $v$.
 The tree is rooted at an arbitrary leaf.
 For each vertex $v$, the minimum weight path $P_v$ visiting
 $v$ and all of $v$'s children (not necessarily starting at $v$) is computed.
 The final tree $T_4$ consists of the union of the paths $\{P_v\}$.
 Again, for the analysis we think of each path $P_v$ replacing
 the edges between $v$ and its children in $T$.

 \begin{figure}[ht]
 \begin{tabAlgorithm}{{\sc Tree-4}$(V,T)$}
 {Find a degree 4 tree of $V$.}
 \algline Root the MST $T$ at a leaf vertex $r$.
 \algline {\bf For} \= each vertex $v \in V$ do
 \algline       \> Compute the shortest path $P_v$ 
                        visiting $v$ and all its children.
 \algline Return $T_4$, the tree formed by the union of the paths $\{P_v\}$.
 \end{tabAlgorithm}
 \caption{Algorithm to find a degree 4 tree.}
 \label{tree4-alg}
 \end{figure}

\begin{lemma}
The algorithm in Fig.~\ref{tree4-alg} returns a degree-4 spanning tree
of the given set of points $V$.
\end{lemma}

\begin{proof}
 A proof by induction shows that $T_4$ is a tree.
 Each vertex $v$ occurs in at most two paths and thus has degree at most four.
\end{proof}

\begin{lemma} \label{deg4-theorem}
Let $v$ be a vertex in an MST $T$ for a set of points in $\Dim{2}$.
Let $P_v$ be the shortest path visiting $\{v\}\ \cup$ child$_T(v)$.
\[ w(P_v) \leq 1.25 \times \sum_{v_i \in \mbox{child}_T(v)} \overline{v v_i}.\]
\end{lemma}

   From the above lemma, each path $P_v$ weighs at most $1.25$ times
   the net weight of the edges it replaces.  Thus,

\begin{theorem}
Let $T$ be a minimum spanning tree of a set of points in $\Dim{2}$.
Let $T_4$ be the spanning tree output by the algorithm in Fig.~\ref{tree4-alg}.
\[ w(T_4) \leq 1.25 \times w(T).\]
\end{theorem}

\noindent{\it Proof of Lemma~\ref{deg4-theorem}. }
The proof is similar to the proof of Lemma~\ref{deg3-theorem}.
As before, we consider cases depending on the number of children of $v$.
The cases when $v$ has no children, one child, or two children are trivial.

\begin{figure}[htbp]
 \begin{center}
\setlength{\unitlength}{0.0085in}
\begingroup\makeatletter\ifx\SetFigFont\undefined
\def\x#1#2#3#4#5#6#7\relax{\def\x{#1#2#3#4#5#6}}%
\expandafter\x\fmtname xxxxxx\relax \def\y{splain}%
\ifx\x\y   
\gdef\SetFigFont#1#2#3{%
  \ifnum #1<17\tiny\else \ifnum #1<20\small\else
  \ifnum #1<24\normalsize\else \ifnum #1<29\large\else
  \ifnum #1<34\Large\else \ifnum #1<41\LARGE\else
     \huge\fi\fi\fi\fi\fi\fi
  \csname #3\endcsname}%
\else
\gdef\SetFigFont#1#2#3{\begingroup
  \count@#1\relax \ifnum 25<\count@\count@25\fi
  \def\x{\endgroup\@setsize\SetFigFont{#2pt}}%
  \expandafter\x
    \csname \romannumeral\the\count@ pt\expandafter\endcsname
    \csname @\romannumeral\the\count@ pt\endcsname
  \csname #3\endcsname}%
\fi
\fi\endgroup
\begin{picture}(630,215)(0,-10)
\put(545,178){\blacken\ellipse{4}{4}}
\put(545,178){\ellipse{4}{4}}
\put(5,41){\blacken\ellipse{4}{4}}
\put(5,41){\ellipse{4}{4}}
\put(125,41){\blacken\ellipse{4}{4}}
\put(125,41){\ellipse{4}{4}}
\put(65,181){\blacken\ellipse{4}{4}}
\put(65,181){\ellipse{4}{4}}
\put(65,99){\blacken\ellipse{4}{4}}
\put(65,99){\ellipse{4}{4}}
\put(605,41){\blacken\ellipse{4}{4}}
\put(605,41){\ellipse{4}{4}}
\put(545,101){\blacken\ellipse{4}{4}}
\put(545,101){\ellipse{4}{4}}
\put(485,41){\blacken\ellipse{4}{4}}
\put(485,41){\ellipse{4}{4}}
\put(325,41){\blacken\ellipse{4}{4}}
\put(325,41){\ellipse{4}{4}}
\put(445,41){\blacken\ellipse{4}{4}}
\put(445,41){\ellipse{4}{4}}
\put(385,101){\blacken\ellipse{4}{4}}
\put(385,101){\ellipse{4}{4}}
\put(385,181){\blacken\ellipse{4}{4}}
\put(385,181){\ellipse{4}{4}}
\put(283,41){\blacken\ellipse{4}{4}}
\put(283,41){\ellipse{4}{4}}
\put(165,41){\blacken\ellipse{4}{4}}
\put(165,41){\ellipse{4}{4}}
\put(225,181){\blacken\ellipse{4}{4}}
\put(225,181){\ellipse{4}{4}}
\put(225,101){\blacken\ellipse{4}{4}}
\put(225,101){\ellipse{4}{4}}
\path(285,41)(225,101)(165,41)(225,181)
\path(325,41)(385,101)(445,41)(385,181)
\path(485,41)(545,101)(545,181)(605,41)
\path(65,181)(65,101)(5,41)(125,41)
\put(60,188){\makebox(0,0)[lb]{\smash{{{\SetFigFont{12}{14.4}{rm}$v_3$}}}}}
\put(75,98){\makebox(0,0)[lb]{\smash{{{\SetFigFont{12}{14.4}{rm}$v$}}}}}
\put(0,18){\makebox(0,0)[lb]{\smash{{{\SetFigFont{12}{14.4}{rm}$v_1$}}}}}
\put(55,3){\makebox(0,0)[lb]{\smash{{{\SetFigFont{12}{14.4}{rm}$P_1$}}}}}
\put(120,18){\makebox(0,0)[lb]{\smash{{{\SetFigFont{12}{14.4}{rm}$v_2$}}}}}
\put(160,18){\makebox(0,0)[lb]{\smash{{{\SetFigFont{12}{14.4}{rm}$v_1$}}}}}
\put(215,3){\makebox(0,0)[lb]{\smash{{{\SetFigFont{12}{14.4}{rm}$P_2$}}}}}
\put(220,108){\makebox(0,0)[lb]{\smash{{{\SetFigFont{12}{14.4}{rm}$v$}}}}}
\put(215,188){\makebox(0,0)[lb]{\smash{{{\SetFigFont{12}{14.4}{rm}$v_3$}}}}}
\put(280,18){\makebox(0,0)[lb]{\smash{{{\SetFigFont{12}{14.4}{rm}$v_2$}}}}}
\put(320,18){\makebox(0,0)[lb]{\smash{{{\SetFigFont{12}{14.4}{rm}$v_1$}}}}}
\put(375,3){\makebox(0,0)[lb]{\smash{{{\SetFigFont{12}{14.4}{rm}$P_3$}}}}}
\put(380,108){\makebox(0,0)[lb]{\smash{{{\SetFigFont{12}{14.4}{rm}$v$}}}}}
\put(440,18){\makebox(0,0)[lb]{\smash{{{\SetFigFont{12}{14.4}{rm}$v_2$}}}}}
\put(480,18){\makebox(0,0)[lb]{\smash{{{\SetFigFont{12}{14.4}{rm}$v_1$}}}}}
\put(535,3){\makebox(0,0)[lb]{\smash{{{\SetFigFont{12}{14.4}{rm}$P_4$}}}}}
\put(600,18){\makebox(0,0)[lb]{\smash{{{\SetFigFont{12}{14.4}{rm}$v_2$}}}}}
\put(535,188){\makebox(0,0)[lb]{\smash{{{\SetFigFont{12}{14.4}{rm}$v_3$}}}}}
\put(375,188){\makebox(0,0)[lb]{\smash{{{\SetFigFont{12}{14.4}{rm}$v_3$}}}}}
\put(540,78){\makebox(0,0)[lb]{\smash{{{\SetFigFont{12}{14.4}{rm}$v$}}}}}
\end{picture}

 \end{center}
 \caption{$T_4$, three children}
 \label{d4-3c}
 \end{figure}
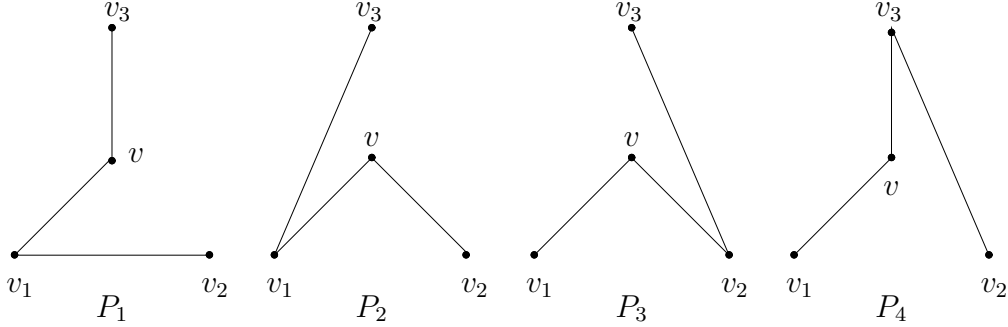

\noindent {\em Case 1: $v$ has 3 children, $v_1,v_2,v_3$. }
Let $v_1$ be the point that is closest to $v$, among its
children. Consider the following four paths (see Fig.~\ref{d4-3c}):
$P_1 = [v_2,v_1,v,v_3], P_2 = [v_2,v,v_1,v_3], 
        P_3 = [v_1,v,v_2,v_3]$ and $P_4 = [v_1,v,v_3,v_2]$.

 Clearly,
   \[ w(P_v) \le  \frac{w(P_1)}{3} + \frac{w(P_2)}{3}
        +\frac{w(P_3)}{6} + \frac{w(P_4)}{6}. \]
We will show that
\[
\frac{w(P_1)}{3} + \frac{w(P_2)}{3} +\frac{w(P_3)}{6} + \frac{w(P_4)}{6}
        \leq \frac{2+\sqrt{3}}{3} (\Edge{vv_1} + \Edge{vv_2} + \Edge{vv_3}).
\]
   This proves the three-child case because $\frac{2+\sqrt{3}}{3}$
approximately equals 1.244 and is less than 1.25. 
   This simplifies to
\[\frac{\Edge{v_1v_2} + \Edge{v_1v_3} + \Edge{v_2v_3}}{3} + \Edge{vv_1} 
        + \frac{\Edge{vv_2} + \Edge{vv_3}}{2}
        \leq \frac{2+\sqrt{3}}{3} (\Edge{vv_1} + \Edge{vv_2} + \Edge{vv_3}),
\]
which further simplifies to
\begin{equation} \label{eq-to-prove}
   \Perimeter{v_1v_2v_3} \leq (\sqrt{3}-1) \Edge{vv_1}
        +(\sqrt{3} + \frac{1}{2}) (\Edge{vv_2} + \Edge{vv_3}).
\end{equation}
Since $v_1$ is the closest point to $v$, applying Lemma~\ref{triangle-lemma},
we get
\[ \Perimeter{v_1v_2v_3} 
        \leq (3\sqrt{3}-4) \Edge{vv_1} + 2 (\Edge{vv_2} + \Edge{vv_3}).
\]
and hence
\begin{eqnarray*}
  \Perimeter{v_1v_2v_3}  &\leq&  (\sqrt{3}-1) \Edge{vv_1}
        + (2 \sqrt{3} - 3) \Edge{vv_1} + 2 (\Edge{vv_2} + \Edge{vv_3}) \\
  &\leq&  (\sqrt{3}-1) \Edge{vv_1} +  (\sqrt{3} + \frac{1}{2})
        (\Edge{vv_2} + \Edge{vv_3}).
\end{eqnarray*}

This proves (\ref{eq-to-prove}).

\noindent {\em Case 2: $v$ has 4 children, $v_1,v_2,v_3,v_4$. }
Assume that $v_1$ is the point that is closest to $v$, among its 
children. Let the  order of the points be
$v_1,v_2,v_3,v_4$, when we scan the plane clockwise from $v$, starting from 
an arbitrary direction.

There are two cases, depending on whether $v_4$ or $v_3$ is the point
that is furthest from $v$ among its children. We first address the
case when $v_4$ is the furthest point.
(The proof for the case when $v_2$ is the point furthest from $v$ 
is symmetric to the case when $v_4$ is the furthest point.)

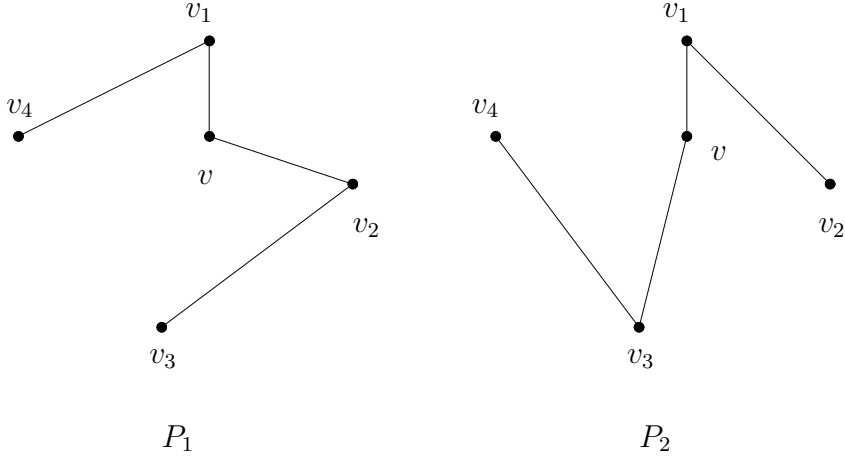
\begin{figure}[htbp]
 \begin{center}
\setlength{\unitlength}{0.01250000in}
\begingroup\makeatletter\ifx\SetFigFont\undefined
\def\x#1#2#3#4#5#6#7\relax{\def\x{#1#2#3#4#5#6}}%
\expandafter\x\fmtname xxxxxx\relax \def\y{splain}%
\ifx\x\y   
\gdef\SetFigFont#1#2#3{%
  \ifnum #1<17\tiny\else \ifnum #1<20\small\else
  \ifnum #1<24\normalsize\else \ifnum #1<29\large\else
  \ifnum #1<34\Large\else \ifnum #1<41\LARGE\else
     \huge\fi\fi\fi\fi\fi\fi
  \csname #3\endcsname}%
\else
\gdef\SetFigFont#1#2#3{\begingroup
  \count@#1\relax \ifnum 25<\count@\count@25\fi
  \def\x{\endgroup\@setsize\SetFigFont{#2pt}}%
  \expandafter\x
    \csname \romannumeral\the\count@ pt\expandafter\endcsname
    \csname @\romannumeral\the\count@ pt\endcsname
  \csname #3\endcsname}%
\fi
\fi\endgroup
\begin{picture}(370,210)(0,-10)
\put(345,113){\blacken\ellipse{4}{4}}
\put(345,113){\ellipse{4}{4}}
\put(285,173){\blacken\ellipse{4}{4}}
\put(285,173){\ellipse{4}{4}}
\put(265,53){\blacken\ellipse{4}{4}}
\put(265,53){\ellipse{4}{4}}
\put(285,133){\blacken\ellipse{4}{4}}
\put(285,133){\ellipse{4}{4}}
\put(205,133){\blacken\ellipse{4}{4}}
\put(205,133){\ellipse{4}{4}}
\put(65,53){\blacken\ellipse{4}{4}}
\put(65,53){\ellipse{4}{4}}
\put(5,133){\blacken\ellipse{4}{4}}
\put(5,133){\ellipse{4}{4}}
\put(145,113){\blacken\ellipse{4}{4}}
\put(145,113){\ellipse{4}{4}}
\put(85,173){\blacken\ellipse{4}{4}}
\put(85,173){\ellipse{4}{4}}
\put(85,133){\blacken\ellipse{4}{4}}
\put(85,133){\ellipse{4}{4}}
\path(5,133)(85,173)(85,133)
        (145,113)(65,53)
\path(205,133)(265,53)(285,133)
        (285,173)(345,113)
\put(65,3){\makebox(0,0)[lb]{\smash{{{\SetFigFont{12}{14.4}{rm}$P_1$}}}}}
\put(60,38){\makebox(0,0)[lb]{\smash{{{\SetFigFont{12}{14.4}{rm}$v_3$}}}}}
\put(0,143){\makebox(0,0)[lb]{\smash{{{\SetFigFont{12}{14.4}{rm}$v_4$}}}}}
\put(80,113){\makebox(0,0)[lb]{\smash{{{\SetFigFont{12}{14.4}{rm}$v$}}}}}
\put(75,183){\makebox(0,0)[lb]{\smash{{{\SetFigFont{12}{14.4}{rm}$v_1$}}}}}
\put(145,93){\makebox(0,0)[lb]{\smash{{{\SetFigFont{12}{14.4}{rm}$v_2$}}}}}
\put(195,143){\makebox(0,0)[lb]{\smash{{{\SetFigFont{12}{14.4}{rm}$v_4$}}}}}
\put(275,183){\makebox(0,0)[lb]{\smash{{{\SetFigFont{12}{14.4}{rm}$v_1$}}}}}
\put(340,93){\makebox(0,0)[lb]{\smash{{{\SetFigFont{12}{14.4}{rm}$v_2$}}}}}
\put(260,38){\makebox(0,0)[lb]{\smash{{{\SetFigFont{12}{14.4}{rm}$v_3$}}}}}
\put(265,3){\makebox(0,0)[lb]{\smash{{{\SetFigFont{12}{14.4}{rm}$P_2$}}}}}
\put(295,123){\makebox(0,0)[lb]{\smash{{{\SetFigFont{12}{14.4}{rm}$v$}}}}}
\end{picture}

 \end{center}
 \caption{$T_4$, four children}
 \label{d4-4c}
 \end{figure}

Consider the following paths:
$P_1 = [v_4, v_1, v, v_2, v_3]$ and $P_2 = [v_4, v_3, v, v_1, v_2]$
(see Fig.~\ref{d4-4c}).

The path $P_v$ added by the algorithm is at most as heavy as the lighter of
the paths $P_1$ and $P_2$. Hence
\[ w(P_v) \le \min (P_1,P_2) \leq \frac{w(P_1)+w(P_2)}{2}. \]

We will show that
\[ \frac{w(P_1)+w(P_2)}{2} \le 
        1.25 (\Edge{vv_1} + \Edge{vv_2} + \Edge{vv_3} + \Edge{vv_4}).
\]

Simplifying, we need to show that
\[ \frac{1}{2}( \Edge{v_4v_1} + \Edge{v_1v} + \Edge{vv_2} + \Edge{v_2v_3} 
        + \Edge{v_4v_3} + \Edge{v_3v} + \Edge{vv_1} + \Edge{v_1v_2})
\leq \frac{5}{4} (\Edge{vv_1} + \Edge{vv_2} + \Edge{vv_3} + \Edge{vv_4}).
\]

Further simplifying, we get:
\[ \Perimeter{v_1 v_2 v_3 v_4} 
        \leq \frac{1}{2} \Edge{vv_1} + \frac{5}{2} \Edge{vv_4} 
                + \frac{3}{2} (\Edge{vv_2} + \Edge{vv_3}).
\]

Note that if it happens that $v_3$ was the farthest point from $v$,
among its children, we get a similar equation with $v_3$ and $v_4$
being exchanged in r.h.s of the equation.  By symmetry, the case when
$v_2$ is furthest is similar to $v_4$ being farthest.

Without loss of generality, 
$\Edge{vv_3}\geq\Edge{vv_2}$. The proof now proceeds in a manner
similar to the proof of Lemma~\ref{triangle-lemma}. If there is a
configuration of points for which this equation is not true (the l.h.s
exceeds the r.h.s) then we can move $v_4, v_3$ closer to $v$ until
$\Edge{vv_2}=\Edge{vv_3}=\Edge{vv_4}$. In doing this, we decrease the
l.h.s by at most $2(\Edge{vv_4}-\Edge{vv_2})+2(\Edge{vv_3}-\Edge{vv_2})$.  
Clearly, the r.h.s decreases by exactly
$4(\Edge{vv_4}-\Edge{vv_2})+4(\Edge{vv_3}-\Edge{vv_2})$. This ensures
that the l.h.s is still greater than the r.h.s. Hence without loss of
generality, if there is a configuration for which our equation is not
true then there is a configuration with the property that
$\Edge{vv_4}=\Edge{vv_3}=\Edge{vv_2}$.  We now show that when this
property is true there is no counter-example.

By scaling, we may assume that $\Edge{vv_4}=\Edge{vv_3}=\Edge{vv_2}=1$,
and $\Edge{vv_1}=\epsilon$, where $\epsilon\leq1$.

Note that (by Corollary \ref{sixty-degree-theorem}) 
  $v$ was originally within the convex hull of its four children.
  Also (by Corollary \ref{convexity-corollary}),
  every child is on the convex hull.
  These properties are both maintained by the above shrinking steps.

\begin{figure}[htbp]
 \begin{center}
\setlength{\unitlength}{0.009in}
\begingroup\makeatletter\ifx\SetFigFont\undefined
\def\x#1#2#3#4#5#6#7\relax{\def\x{#1#2#3#4#5#6}}%
\expandafter\x\fmtname xxxxxx\relax \def\y{splain}%
\ifx\x\y   
\gdef\SetFigFont#1#2#3{%
  \ifnum #1<17\tiny\else \ifnum #1<20\small\else
  \ifnum #1<24\normalsize\else \ifnum #1<29\large\else
  \ifnum #1<34\Large\else \ifnum #1<41\LARGE\else
     \huge\fi\fi\fi\fi\fi\fi
  \csname #3\endcsname}%
\else
\gdef\SetFigFont#1#2#3{\begingroup
  \count@#1\relax \ifnum 25<\count@\count@25\fi
  \def\x{\endgroup\@setsize\SetFigFont{#2pt}}%
  \expandafter\x
    \csname \romannumeral\the\count@ pt\expandafter\endcsname
    \csname @\romannumeral\the\count@ pt\endcsname
  \csname #3\endcsname}%
\fi
\fi\endgroup
\begin{picture}(296,263)(0,-10)
\put(142.211,153.105){\arc{15.734}{3.9885}{6.4241}}
\put(139.263,141.368){\arc{23.182}{5.4432}{8.0505}}
\put(136,146){\blacken\ellipse{4}{4}}
\put(136,146){\ellipse{4}{4}}
\put(256,206){\blacken\ellipse{4}{4}}
\put(256,206){\ellipse{4}{4}}
\put(16,206){\blacken\ellipse{4}{4}}
\put(16,206){\ellipse{4}{4}}
\put(136,11){\blacken\ellipse{4}{4}}
\put(136,11){\ellipse{4}{4}}
\put(136,226){\blacken\ellipse{4}{4}}
\put(136,226){\ellipse{4}{4}}
\path(136,226)(256,206)(136,11)
        (16,206)(136,226)
\dashline{4.000}(136,226)(136,11)
\dashline{4.000}(16,206)(136,146)
\dashline{4.000}(136,146)(256,206)
\put(0,217){\makebox(0,0)[lb]{\smash{{{\SetFigFont{12}{14.4}{rm}$v_4$}}}}}
\put(126,236){\makebox(0,0)[lb]{\smash{{{\SetFigFont{12}{14.4}{rm}$v_1$}}}}}
\put(122,190){\makebox(0,0)[lb]{\smash{{{\SetFigFont{12}{14.4}{rm}$\epsilon$}}}}}
\put(156,126){\makebox(0,0)[lb]{\smash{{{\SetFigFont{12}{14.4}{rm}$\alpha$}}}}}
\put(116,135){\makebox(0,0)[lb]{\smash{{{\SetFigFont{12}{14.4}{rm}$v$}}}}}
\put(145,3){\makebox(0,0)[lb]{\smash{{{\SetFigFont{12}{14.4}{rm}$v_3$}}}}}
\put(146,168){\makebox(0,0)[lb]{\smash{{{\SetFigFont{12}{14.4}{rm}$\beta$}}}}}
\put(266,201){\makebox(0,0)[lb]{\smash{{{\SetFigFont{12}{14.4}{rm}$v_2$}}}}}
\end{picture}

 \end{center}
 \caption{Figure to illustrate degree four case.}
 \label{deg4}
 \end{figure}
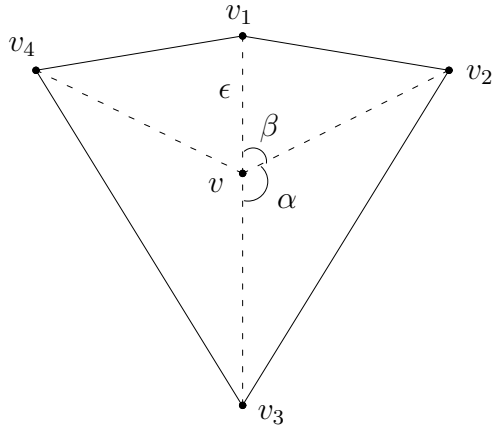

We now wish to prove that 
\[ 
\Perimeter{v_1v_2v_3v_4} \leq \frac{11}{2} + \frac{1}{2} \epsilon.
\]
It is easily shown using elementary calculus that for any $\epsilon$
such that $v_1$ is on the convex hull of the points $\{v_1,\ldots,v_4\}$,
rotating $v_1$ and $v_3$ around $v$ until 
$\Angle{v_1vv_2}=\Angle{v_1vv_4}$ (see Fig.~\ref{deg4}) and
$\Angle{v_2vv_3}=\Angle{v_4vv_3}$ does not decrease the perimeter.
Also, it maintains that $v_1$ is on the convex hull.
Assume the two pairs of angles are equal, and define
$F(\epsilon) = \Perimeter{v_1v_2v_3v_4} - \epsilon/2 - 11/2$.
We will show $F$ is non-positive over the domain of possible~$\epsilon$.

As a function of $\epsilon$, function $F$ is a sum of convex functions
minus a linear function, and thus is convex.  Therefore, $F$ is maximized
either when $\Edge{vv_1} = 1$ or when 
$v_1$ is the midpoint of edge $\Edge{v_2v_4}$
(since $v_1$ is on the convex hull, $v_1$ can not cross the edge,
hence this interval contains all possible values for $\epsilon$).

In the first case, all four points lie on a unit circle with center at $v$.
For any four such points, it is easily proven using calculus that
$\Perimeter{v_1v_2v_3v_4}$ is maximized 
when the four points are the vertices of a square
at $4\sqrt{2} \approx 5.66$.  Thus, $F(1) < 0$.

In the second case, $\Perimeter{v_1v_2v_3v_4}=\Perimeter{v_2v_3v_4}$.
As noted previously, this is at most $3\sqrt{3} \approx 5.2$.
Thus, \hbox{$F(\epsilon) < 0$}.

We now deal with the case when $v_3$ is the furthest point.
In this case we take the paths
$P_1 = [v_4, v_1, v, v_2, v_3]$ and $P_2 = [v_3, v_4, v, v_1, v_2]$.
The path $P$ added by the algorithm is at most as heavy as the lighter of
the paths $P_1$ and $P_2$. Hence,
\[ w(P) \le \min (P_1,P_2) \leq \frac{w(P_1)+w(P_2)}{2}. \]
Simplifying, we get
\[ \Perimeter{v_1 v_2 v_3 v_4} 
        \leq \frac{1}{2} \Edge{vv_1} + \frac{5}{2} \Edge{vv_3} 
                + \frac{3}{2} (\Edge{vv_2} + \Edge{vv_4}).
\]
The proof of this is identical to the proof of the previous case.
\myendproof

\section{Points in higher dimensions}

We show how to compute a degree-3 tree ($T_3$) when the
points are in arbitrary dimension $d \geq3$. 
The algorithm for computing the tree is similar   to the algorithm for
computing degree three trees in the plane --- the tree $T_3$ is formed
by rooting the MST and taking the union of the paths $\{P_v\}$, 
where each $P_v$ is the shortest path starting at $v$
and visiting all of the children of $v$ in the rooted MST.
It is known that any Euclidean MST has constant 
degree \cite{RoS} (for any fixed dimension),
so that the algorithm still requires only linear time.
The bound on the weight of $T_3$ is similar, except that $v$ may have
more children.  We prove that regardless of the number of children
that $v$ has, the weight of $P_v$ is at most $5/3$ the weight 
of the edges that it replaces:

\begin{lemma} \label{3d-lemma}
Let $\{v, v_1, v_2, \ldots, v_k\}$ be a set of arbitrary points in $\DimD$.
There is a  path $P$, starting at $v$, that visits 
all the points $v_1, v_2, \ldots, v_k$ such that
\[ w(P) \leq \frac{5}{3} \sum_{i=1}^{k} \Edge{vv_i}.\]
\end{lemma}

\begin{proof}
We prove this by induction on the degree of $v$.
Sort the points in increasing distance from $v$ as $v_1, \ldots, v_k$.
Let $v=v_0$.
   The lemma is trivially true when $k=0,1,2$. 
Let us assume that the lemma is true for all values of $k$ up to some
$\ell\ge 2$. Consider $k=\ell+1$.
By the induction hypothesis, the claim is true  when $v$ has $k-3$ children;
hence we can find a path $P'$ that starts at $v$ and visits all vertices
$v_i \ (i=1, \ldots, k-3)$ (not necessarily in that order)
such that $w(P') \leq \frac{5}{3} \sum_{i=1}^{k-3} \Edge{v v_i}$.
Let $v_j$ be the last vertex on the path $P'$.
We  add the cheapest path $P''$ that starts
at $v_j$ and visits $v_{k-2}, v_{k-1}$ and $v_k$ (again,
not necessarily in that order).
This path together with $P'$ will form a  path that starts at
$v$ and visits all vertices adjacent to $v$. We  now show that
\begin{equation}
w(P'') \leq \frac{5}{3} (\Edge{vv_{k-2}} + \Edge{vv_{k-1}} + \Edge{vv_k}).
\label{eq-5/3}
\end{equation}
  This suffices to prove the lemma.
   Let $P_1, \ldots, P_6$ be the six possibilities for $P''$.  Clearly,
   \[ w(P'') \leq \frac{1}{6}  \sum_{i=1}^{6} w(P_i).\]
We will prove that 
\[ \frac{1}{6}  \sum_{i=1}^{6} w(P_i) \leq 
        \frac{5}{3} (\Edge{vv_{k-2}} + \Edge{vv_{k-1}} + \Edge{vv_k}) .\]
   This simplifies to
\begin{equation} \label{eq-tet+tri}
2\ \Perimeter{v_{k-2} v_{k-1} v_k} + \sum_{i=k-2}^{k} \Edge{v_{j}v_i}
        \leq 5 (\Edge{vv_{k-2}} + \Edge{vv_{k-1}} + \Edge{vv_k}) .
\end{equation}

Notice that if the above equation is not true, we
can ``shrink'' all the $v_i\  (i = k-2, k-1, k)$ until 
$\Edge{vv_{j}} = \Edge{vv_{k-2}} = \Edge{v v_{k-1}} = \Edge{v v_{k}}$. 
Assume that $\delta = (\Edge{vv_{k-2}} - \Edge{vv_{j}}) +
(\Edge{vv_{k-1}} - \Edge{vv_{j}}) + (\Edge{vv_{k}} - \Edge{vv_{j}})$.
This can be done because the
r.h.s decreases by $5 \delta$, and the l.h.s decreases by at most 
$5 \delta$. If the above equation is not true then it is also not
true when the distance from $v$ to all the points is the same. By
scaling, we can assume that the distance of the points from $v$ is~1.
We call this a canonical configuration.
The following proposition is implied by Lillington's work \cite{Li} 
and helps in completing the proof.

\begin{proposition} \label{tetrahedron-lemma}
Let $A,B,C$ and $D$ be points on a unit sphere in $d$-dimensions, $d\ge3$.
The function
$F=\Edge{AB}+\Edge{AC}+\Edge{AD}+\Edge{BC}+\Edge{CD}+\Edge{BD}$
reaches a maximum value of $4\sqrt{6}$ when the points $A,B,C$ and $D$
form a regular tetrahedron.
\end{proposition}

We will now show that (\ref{eq-tet+tri}) is satisfied by the canonical
configuration.
The left side of (\ref{eq-tet+tri}) can be written as the sum of the
sides of the tetrahedron formed by the points
$\{v_k,v_{k-1},v_{k-2},v_{j}\}$ and the sum of the sides
of the triangle formed by the points $\{v_k,v_{k-1},v_{k-2}\}$.
These points lie on a sphere whose center is $v$. 
By Lemma~\ref{tetrahedron-lemma},
the first sum is bounded by $4\sqrt{6}$.
The second sum is bounded by $3\sqrt{3}$.
Hence the left side of (\ref{eq-tet+tri}) is bounded by 
$4\sqrt{6}+3\sqrt{3}$, which is about 14.994.
The right side of (\ref{eq-tet+tri}) is 15.
Hence (\ref{eq-tet+tri}) is satisfied by the canonical configuration
and therefore all configurations.
This concludes the proof of Lemma~\ref{3d-lemma}.
\end{proof}

\noindent{\bf Remark.}
The algorithm outlined earlier runs in linear time only when $d$, the
number of dimensions, is a constant.
The algorithm can be modified to run in linear time for all $d$ as follows.
Observe that in the proof of Lemma~\ref{3d-lemma}, we considered the
neighbors of $v$ only three at a time.
Therefore the algorithm could also group vertices into sets of 3 each,
based on the distance from $v$, and inductively construct the path
as in the proof of the lemma.
This algorithm would have the same performance guarantee (5/3) as the
earlier algorithm for constructing a degree-3 tree, and in addition
have the added advantage of running in linear time for all dimensions.

\section{Conclusions} \label{sec-conc}
We have given a simple algorithm for computing a degree-3 (degree-4)
tree for points in the plane that is  within 1.5 (1.25) 
of an MST of the points.
An extension of the algorithm finds a degree-3 tree of an arbitrary
set of points in $d$-dimensions within 5/3 of an MST.
If an MST of the points is given as part of the input,
our algorithms run in linear time.
All our proofs are based on elementary geometric techniques.

Though our algorithms improve greatly the best known ratios for each
of the respective problems, there are still large gaps between the
ratios that we obtain and the best  bounds that we think are achievable.
For example, in the case of points in the plane, consider the ratio of
the weight of a minimum weight degree-3 tree to the weight of an MST.
The worst example that we can obtain for this ratio is
$\frac{\sqrt{2}+3}{4} \approx 1.104$
(with 5 points, where 4 of the points are at the corners
of a square and the fifth point is in the middle).
There is a large gap between this and the ratio of 1.5 obtained by our 
algorithm. Is 1.104 the worst case ratio?
Are there polynomial time algorithms which obtain factors better than 1.5?
Notice that the performance ratio obtained by our algorithm on the
example in Fig.~\ref{bad-ex} is highly sensitive to the vertex 
chosen as the root. One potential algorithm is to simply try all possible
vertices as the root, and to pick the tree of minimum weight.
Does such an algorithm have a better performance guarantee?

For the problem of finding degree-4 trees, our algorithm obtains a
ratio of 1.25.  Unlike degree-3 trees, we are unable to show that this
ratio is tight for the algorithm. 
Can the factor of 1.25 for the algorithm be improved?
The worst example for the ratio between a minimum-weight degree-4 tree
and an MST that we can obtain is about 1.035 (5 points on the vertices
of a regular pentagon with a sixth point in their centroid).
Are there examples with worse ratios?

Problems of approximating degree-$k$ trees in higher dimensions and in
general metric spaces within factors better than 2 are still open.

\section*{Acknowledgments}
We thank Andras Bezdek for telling us about \cite{Li}. We thank
Karoly Bezdek and Bob Connelly for useful discussions
and the committee members of STOC '94 for simplifying
the proof of Lemma~\ref{triangle-lemma} and for pointing out \cite{MS}.

\end{document}